\documentclass[proof]{rmaa}

\usepackage{paralist}
\usepackage{psfrag,color}
\usepackage{rotating}
\usepackage[latin1]{inputenc}

\newcommand{\hda}{HD\,108}
\newcommand{\hdc}{HD\,148937}
\newcommand{\hdb}{HD\,191612}
\newcommand{\thet}{$\theta^1$~Ori~C}
\newcommand{\lo}{$\lambda$}
\newcommand{\ha}{H$\alpha$}
\newcommand{\hb}{H$\beta$}
\newcommand{\hei}{He\,{\sc i}}
\newcommand{\heii}{He\,{\sc ii}}
\newcommand{\niii}{N\,{\sc{iii}}}
\newcommand{\ciii}{C\,{\sc{iii}}}
\newcommand{\civ}{C\,{\sc{iv}}}
\newcommand{\oiii}{O\,{\sc{iii}}}
\newcommand{\kms}{km\,s$^{-1}$}

\newcommand{\msol}{M$_{\odot}$}
\newcommand{\xmm}{{\sc XMM}\emph{-Newton}}

\title{The mysterious Of?p class and the magnetic O-star \thet: confronting observations} 

\author{Ya\"el Naz\'e\altaffilmark{1} Nolan R. Walborn\altaffilmark{2} and Fabrice Martins\altaffilmark{3}}
\altaffiltext{1}{Postdoctoral Researcher FNRS; Institut d'Astrophysique et de G\'eophysique, Universit\'e de Li\`ege, 
All\'ee du 6 Ao\^ut 17, Bat B5C, B4000-Li\`ege, Belgium}
\altaffiltext{2}{Space Telescope Science Institute, 3700 San Martin Drive, Baltimore, MD 21218, USA}
\altaffiltext{3}{Universit\'e Montpellier II - GRAAL, CNRS - UMR 5024, place Eug\`ene Bataillon, F-34095 Montpellier, France}

\shortauthor{Naz\'e, Walborn, \& Martins}
\shorttitle{Confronting observations of O${\rm f?p}$ stars and \thet}

\listofauthors{Y. Naz\'e, N.R. Walborn \& F. Martins}
\indexauthor{Naz\'e, Y.}
\indexauthor{Walborn, N.R.}
\indexauthor{Martins, F.}

\SetVolume{00}
\SetYear{2008}

\ReceivedDate{}
\AcceptedDate{}

\abstract{In recent years, the stars of the Of?p category have revealed a 
wealth of peculiar phenomena: varying line profiles, photometric changes, 
and X-ray overluminosity are only a few of their characteristics. Here we 
review their physical properties, to facilitate comparisons among the 
Galactic members of this class. As one of them has been proposed to resemble 
the magnetic oblique rotator \thet, though with a longer period, this latter 
object is also included in our study to illuminate its similarities and 
differences with the Of?p category.}
\resumen{En a\~nos recientes, las estrellas de la categor\'{\i}a Of?p han
revelado una multitud de fen\'omenos peculiares: perfiles de l\'{\i}neas
variables, cambios fotom\'etricos, y exceso de luminosidad en rayos X son
solamente algunas de sus caracter\'{\i}sticas.  Aqu\'{\i} repasamos sus
propiedades f\'{\i}sicas, para facilitar comparaciones entre los miembros
Gal\'acticos de esta clase.  Como uno de ellos se ha propuesto como
semejante al rotador magn\'etico oblicuo \thet, aunque con un per\'{\i}odo
m\'as largo, este \'ultimo objeto est\'a tambi\'en inclu\'{\i}do en nuestro
estudio, para iluminar sus similitudes y diferencias con la categor\'{\i}a 
Of?p.}

\addkeyword{Stars: early-type}
\addkeyword{Stars: individual (HD\,108, HD\,148937, HD\,191612, $\theta^1$~Ori~C)}
\addkeyword{Stars: variables: other}

\begin{document}
\maketitle

\section{Introduction}

In astronomy, the study of peculiar objects often provides crucial 
information.  Indeed, a rare phenomenon could be linked to short-lived 
phases in the evolution of stars or galaxies. It might thus represent 
a `missing link' in our understanding of the Universe, therefore deserving 
an in-depth analysis. 

Though rare, O-type stars are nevertheless important as they are the main
sources of ionizing radiation, mechanical energy, and certain chemical
elements in galaxies. However, their lifetimes are short (a few million 
years) and their rapid evolutionary phases are thus very difficult to 
study. To better understand these hot, massive objects, it is thus important
to probe the properties of the most peculiar ones, such as the Of?p stars.

The Of?p category was defined by \citet{wal72,wal73} to classify stellar 
spectra displaying certain peculiarities, notably strong \ciii\ 
emission lines around 4650\,\AA\ (more details
on classification criteria below). At first, only two stars belonged to 
this category, \hda\ and \hdc, but a third one was soon added, \hdb\ 
\citep{wal73}. A few others have now been identified in the Magellanic Clouds 
\citep{hey92,wal00,mas01,eva04}. In recent years, these peculiar objects 
attracted quite a lot of attention because of the discovery of recurrent 
line-profile variations and X-ray overluminosities. A magnetic field was 
even reported for \hdb, making it the second O-type star with a detected 
magnetic signature - the first one being \thet\ \citep{don02,don06}. 

The aim of this short review is to summarize the properties of the Of?p 
stars, in the visible, UV, and X-ray domains, and to compare them to those of 
\thet, a known magnetic oblique rotator. This paper is organized as follows. 
Section 2 describes the main properties derived from the UV/visible spectra of 
these stars, Section 3 presents their photometric characteristics, Section 4 
describes their observations in the X-ray domain, and Section 5 gives our 
conclusions.

\section{Visible and UV Spectrum} 

The study of O-type stars has long focused on the UV/visible spectra of these 
objects. Indeed, a wealth of information can be derived from them. To cite 
only a few: lines varying in position indicate binarity; relative line 
strengths indicate temperature; and emission-line profile variations 
relate to peculiar phenomena such as transient wind inhomogeneities, 
wind-wind collisions, or a disk confined by an oblique magnetic field.
We present here the properties derived from the spectra of the 
Of?p stars and of \thet.

\subsection{Spectral Type}
Though \thet\ is often studied, its spectral classification is still
under debate. It is generally considered to be O7~V \citep{sta08}, but 
\citet{wal81} found a changing spectral type, which remains to be confirmed. 
In the context of this review, it is important to note the absence of strong 
\ciii\,\lo\,4650 in emission, which is one of the main characteristics
defining Of?p stars.

The f?p designation should only be added when several spectral 
criteria are fulfilled. The most important one is of course the presence 
of a strong \ciii\,\lo\,4650 emission, i.e.\ of an intensity comparable 
to that of the neighboring \niii\ lines. Additional features are common 
amongst Galactic Of?p stars: composite profiles for the Balmer hydrogen 
lines, with a narrow P\,Cygni/emission component superimposed on a broad 
(photospheric) absorption; P\,Cygni \hei\ profile or asymmetric \hei\ 
lines, with the red wing being steeper than the blue one; other peculiarities
such as  weak Si\,{\sc{iv}}\,\lo\lo\,1393,1402 wind profiles unlike 
for Of supergiants, and, in some cases, the 
Si\,{\sc{iii}}\,\lo\lo\,4552,4568,4575 triplet in emission. 

Comparing the \hei\ to \heii\ absorption lines, \hdc\ appears rather early, 
with a type O5.5-6f?p \citep{naz08}, but the cases of \hda\ and \hdb\ seem 
more complex. In fact, the spectral types of these two stars are not constant: 
using the classical \hei-\heii\ indicators, they alternate between O4 
and O8.5 for \hda\ \citep{naz01,naz04b}, or O6.5 and O8 for \hdb\ 
\citep{how07}. As the \ciii\,\lo\,4650 emission is only present when
these stars appear very early, the f?p designation is only added in this 
case, whereas the fp classification is sufficient when the spectral type 
is the latest\footnote{It is interesting to note an additional peculiarity
when the spectral type is the latest: strong \heii\,\lo\,4686 emission 
without the Si\,{\sc{iv}} absorption typical of supergiants.}.

\subsection{Line Profile Variability}
Hints at spectral changes in \thet\ were already reported by \citet{con72}. 
The variability was further studied by \citet{wal81, sta93, wal94, sta96, 
sta08}. The largest variations are observed for the Balmer hydrogen and 
\heii\,\lo\,4686 lines (Fig. \ref{fig:thetvar}). However, the equivalent 
widths (EWs) of other lines, e.g. the photospheric lines of \hei\,\lo\,4471, 
\heii\,\lo\,4542, \oiii\,\lo\,5592, \civ\,\lo\,5801, appear to change in 
harmony with those of H and \heii\,\lo\,4686, though with a smaller amplitude 
(15\% change vs 100\%) and with an opposite behaviour (maximum $absorption$ 
when maximum $emission$ is seen for H lines). Clear line profile 
variability was also found in the UV range for the Si\,{\sc{iv}}\,\lo\lo\,1393,1402, C\,{\sc iv}\,\lo\lo\,1548,1550 O\,{\sc{v}}\,\lo\,1371 N\,{\sc iv}\,\lo\,1718 lines, which display a maximum 
absorption when the H lines present a maximum emission \citep{wal94,sta96}.
In addition, \citet{wal81} mentioned variations of the 
\hei\,\lo\,4471/\heii\,\lo\,4542 ratio, but this was not observed in 
subsequent studies \citep[see e.g.][]{sta96,sta08}.

\begin{figure}
  \centering
  \caption{Variations over the 15d period of the spectra of \thet\ (left, figure from \citealt{sta96}) and evolution of the red spectrum of \hda\ between 1997 and 2003 (right, figure from  \citealt{naz04b}).}
  \label{fig:thetvar}
\end{figure}

Among the Of?p stars, \hda\ has been the most studied, but conflicting 
results about its nature were reported: a short-period binary 
(\citealt{hut75, asl89}, the system having even survived a supernova 
event according to \citealt{bek76}); a single star experiencing wind 
variability and/or harboring a disc and jets \citep{vre79, und94}; or a 
long-period (a few years) binary \citep{bar99}... To attempt to resolve 
the confusing situation, a 30-year spectroscopic campaign was undertaken 
at the Haute-Provence Observatory \citep{naz01}. A detailed analysis of 
the data allowed most of the older models to be discarded and rather 
unveiled a peculiar phenomemon: dramatic line-profile variations on a 
timescale of decades (Figs. \ref{fig:thetvar} and \ref{fig:ewhd108};
for the first hints of the phenomenon see also \citealt{and73}). The 
Balmer hydrogen and \hei\ lines change from strong P\,Cygni profiles 
to pure absorptions with asymmetric profiles. A few other emission lines 
were also found to be strongly variable (\ciii\,\lo\,4650, \heii\,\lo\,4686). 

\citet{wal03} investigated another Of?p star, \hdb, and discovered 
similar line profile variations. The changes apparently come from a 
varying narrow emission, slightly redshifted, superimposed on the `normal' 
stellar lines. It should, however, be noted that not all lines are varying: 
\heii\,\lo\,4200,4542, \oiii\,\lo\,5592, \ciii\,\lo\,5696, 
\civ\,\lo\,5812,... (all most probably of photospheric origin) appear 
relatively constant in strength \citep{naz01,how07}. Therefore, since 
variability affects \hei\,\lo\,4471 but not \heii\,\lo\,4542, apparent 
spectral-type variations, measured from the \hei/\heii\ ratio, are detected, 
as noted above. Note that these changes have a long-term character, but 
small-scale, short-term variations are also observed for \hdb\ as well 
as \hda\ \citep{how07,naz08}. In the remainder of the paper, we define 
the `quiescent' state as the time when the contamination by excess 
emission is minimum, i.e. when the spectral type appears to be the 
latest.

\begin{figure}
\begin{minipage}{7cm}
  \caption{Evolution of the equivalent widths of \hei\,\lo\,4471
  (crosses), \heii\,\lo\,4542 (filled circles), and \hb\ (open triangles)
  in the spectrum of \hda\ (figure from \citealt{naz06}).}
  \label{fig:ewhd108}
\end{minipage}
\begin{minipage}{9cm}
  \centering
  \caption{Phase-locked variations in \thet\ of the equivalent width of \ha. 
(figure from \citealt{sta96}).}
  \label{fig:ewtheta}
\end{minipage}
\end{figure}

\setcounter{figure}{2}
\begin{figure}
  \centering
  \caption{ Continued. Phase-locked variations in \thet\ of the equivalent width of \heii\,\lo\,4686 (left).
  On the right are shown the normalized and phase-averaged variations of
  the equivalent widths of \hei\,\lo\,4471, \heii\,\lo\,4542, \oiii\,\lo\,5592, and \civ\,\lo\,5801 (figures from \citealt{sta96}).}
\end{figure}

Only little was known about \hdc\ until recently, when we undertook a 
spectroscopic campaign with the SMARTS program at the CTIO 1.5m telescope. 
This monitoring revealed low-level variability in the Balmer and 
\heii\,\lo\,4686 lines, but constancy for \hei, \ciii\,\lo\,4650, and 
other lines \citep{naz08}. Note that if this variability were similar 
in nature to that of \hda\ and \hdb, though with a much lower intensity, 
we would expect no detectable variations of the \hei\ and \ciii, as 
the hydrogen changes are the largest of all lines for these stars.

Because of the lack of high-quality data, the UV spectrum of the 
Of?p stars was only little investigated up to now - though the few IUE 
data available, shown in figure 1 of \citet{wal03}, were of course used 
for model atmosphere fits (see Section 2.4). The only 
significant result is the lack of strong variability between the two IUE 
spectra corresponding at the two extreme states of \hdb\ \citep[data
unfortunately taken with two different spectral resolutions, see][]{how07}. 
However, since these stars are highly variable, one still needs to be 
cautious before drawing conclusions from such a limited dataset. Additional 
observations are thus needed to undertake an in-depth analysis, similar 
to what has been done for \thet.

\subsubsection{Periodicity}
The EW changes of \thet\ are recurrent with a very stable period of 
15.424$\pm$0.001d \citep[Fig. \ref{fig:ewtheta}, see also][]{sta08}, 
while those of \hdb\ display a period of 537.6$\pm$0.4d \citep{wal04,how07}. 
Comparing all the available data, the changes observed in \hda\ also appear 
repeatable, on a timescale of a few decades (approximately 50--60 years, see 
\citealt{naz01,naz06})\footnote{It has sometimes been suggested that, 
despite the numerous similarities between \hdb\ and \hda, the former object 
was much closer in nature to \thet\ than to \hda, based on the `similar'
values of their periods. However, it must be underlined that the period 
ratios are actually comparable, $P(HD\,191612)/P(HD\,108) \sim 
P(\theta^1~Ori~C)/P(HD\,191612)$, indicating that no significant conclusion
can be drawn from such `numerology'.}. Finally, the line profile variations 
of \ha\ in \hdc\ present a possible periodicity of 7.031$\pm$0.003d 
\citep{naz08}. 

\subsection{Binarity}
Long-term radial velocity changes, a typical signature of binary motion, were reported for \thet\ by \citet[see also summary in \citealt{ode01}]{sta93}
and for \hdb\ by \citet{naz07}. Subsequent studies have provided more 
detailed orbital parameters. \thet\ is actually a visual binary and a first 
analysis of interferometric measurements yields a period of 10.9 yr, an 
eccentricity of 0.9, and a mass ratio of 0.45$\pm$0.15, suggesting the 
component spectral types to be O5.5~+~O9.5, and the masses 34~+~15.5~\msol\ 
\citep{krau07}. These parameters were recently revised by \citet{pat08}
who proposed a longer period (about 26 yrs) and a much smaller eccentricity ($e=0.16\pm0.14$) but the authors caution that the results are still 
preliminary since less than half the orbit has actually been observed. Radial 
velocities analyzed by \citet{sta08} are consistent with the interferometric 
measurements, but do not permit to discriminate between both solutions. 
It should also be noted that rapid radial velocity changes are also detected
but they seem uncorrelated with the 15d timescale of the line profile 
variability \citep{sta08}. On the other hand, the radial velocity curve of 
\hdb\ shows that $P$=1542$\pm$14d, $e$=0.44$\pm$0.04, and 
$M_2/M_1$=0.48$\pm$0.04---the system thus probably contains an 
O8~+~early-B star (30~+~15~\msol, \citealt{how07}).

The multiplicity of the other two Of?p objects is less clear. Though \hda\ 
was repeatedly suggested to be a binary in the past \citep{hut75,asl89,bar99},
recent monitoring could discard the proposed solutions and all periods 
from a few days to a few years\citep{naz01}. However, radial-velocity 
changes are detected for \hda\ (between $-$55 and $-$85\,\kms, 
\citealt{naz04,naz06}), and a very long-term binary cannot be 
excluded. \hdc\ displays no radial-velocity changes of amplitude larger 
than 10~\kms\ on short or long timescales \citep{con77,gar80,naz08}, but 
low-amplitude variations or very long-term changes are not excluded by the 
available data.

\subsection{Physical Parameters}
The spectrum of \thet\ was modeled in detail by \citet{sim06} and
\citet{krau07}; their results are reproduced in Table \ref{tab:physpar},
with only one modification. In fact, the distance to the Orion Nebula 
was recently revised by \citet{men07} to 414\,pc (instead of the usual 
450\,pc, used notably by \citealt{sim06}, or 434\,pc of \citealt{krau07}). 
The luminosity of \thet\ appearing in Table \ref{tab:physpar} was thus 
changed accordingly, as well as the stellar radius, assuming the effective
temperature is not changed, which is a fair approximation.

\begin{figure}
  \centering
  \includegraphics[width=8.3cm]{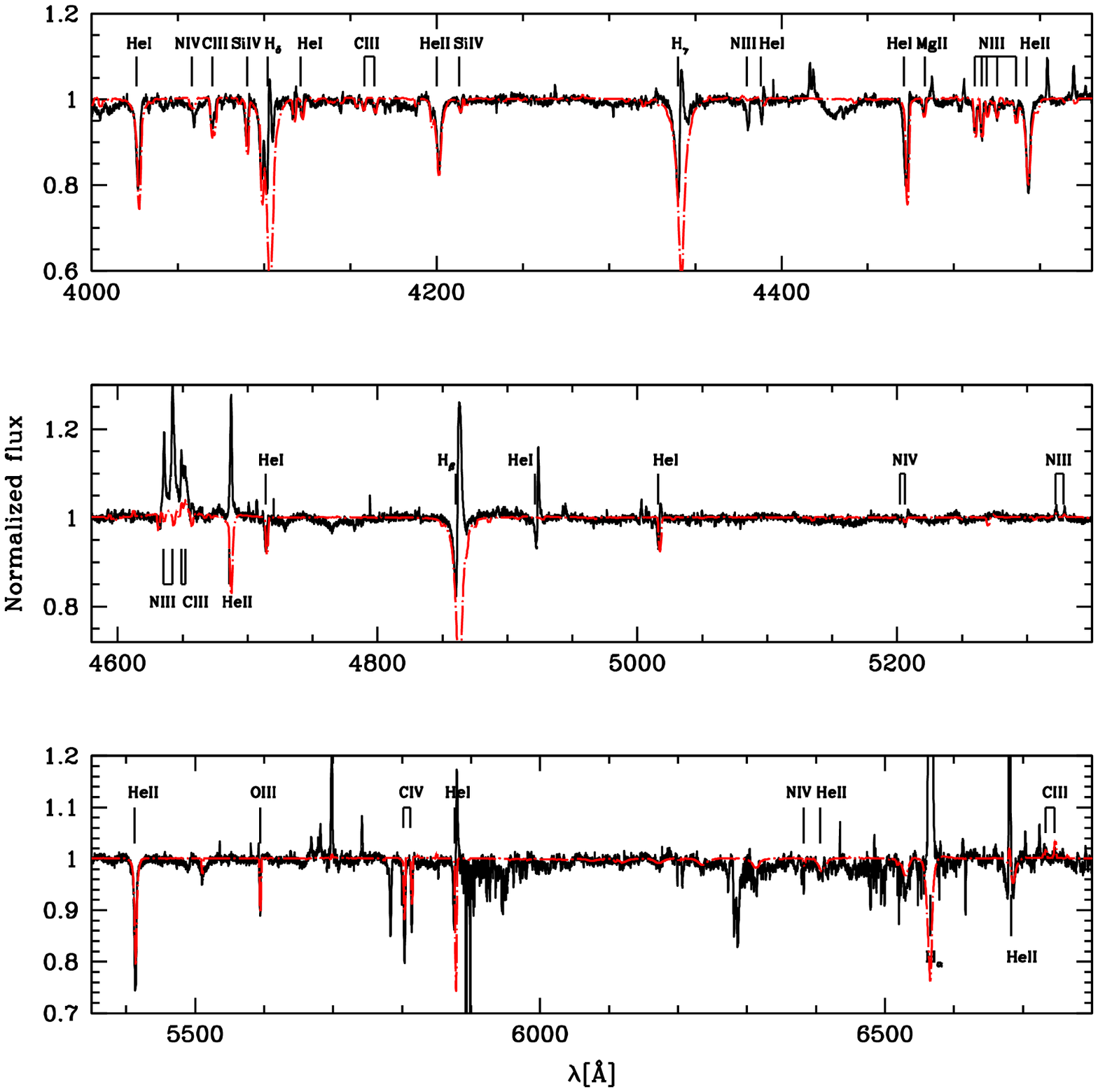}
  \includegraphics[width=8.3cm]{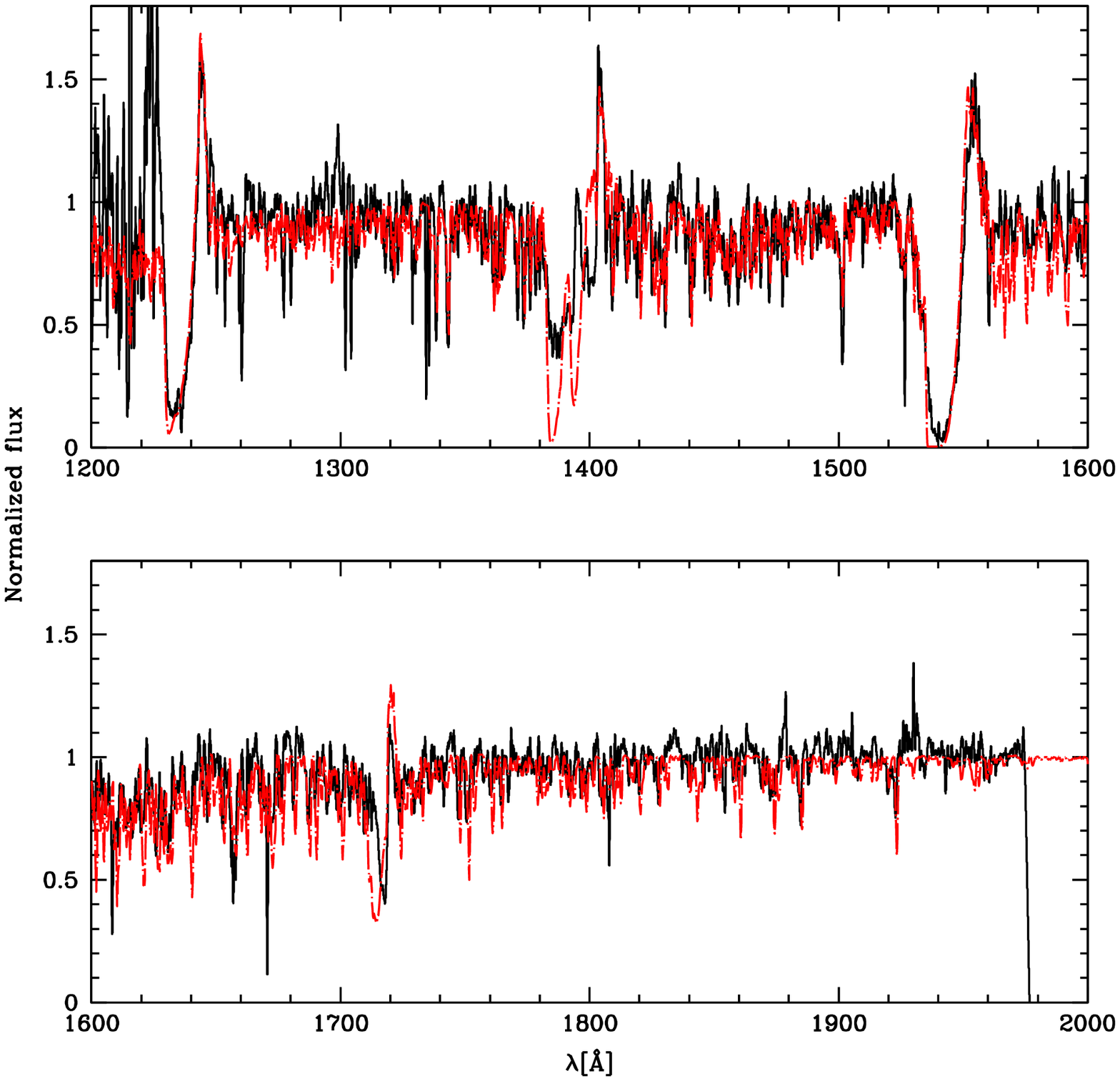}
  \caption{Comparison of the optical and UV spectrum of \hda\ (solid black line) and the best fit CMFGEN model (dash-dot red line, see Table \ref{tab:physpar} for parameters; note that the fit also yields $\dot M$=\,1--3$\times10^{-7}$\,M$_{\odot}$\,yr$^{-1}$ for a clumping factor $f$=0.01). The narrow emission components in the \hei\ and H optical lines are not fitted and are most probably of circumstellar origin (disk?). This figure only appears in colors in the electronic version of the journal. \label{cmfgen}.}
\end{figure}

We estimated the projected rotational velocity of the Of?p objects by 
applying the Fourier method \citep[see][and references therein]{sim07} 
to uncontaminated, `photospheric' lines (e.g. \civ\,\lo\,5812 and 
\oiii\,\lo\,5592). This method was also the one chosen by \citet{sim06}, 
thus enabling a direct comparison between the stars. It might be noted 
that these new values are lower than previous estimates because the 
Fourier method can disentangle the different contributions to the line 
broadening \citep{sim07}. Model-atmosphere fits further provided the
effective temperatures, gravities, and luminosities of the Of?p stars 
in the quiescent state (\citealt{how07} for \hdb, \citealt{naz08} 
for \hdc, and this work for \hda, see also Fig. \ref{cmfgen}). The results 
are reported in Table \ref{tab:physpar}). The best-fit gravity values clearly 
suggest that these stars are not supergiants, as was once proposed, but 
giants or main-sequence objects. 

\begin{figure}
  \centering
\begin{minipage}{8cm}
  \centering
  \includegraphics[width=7cm]{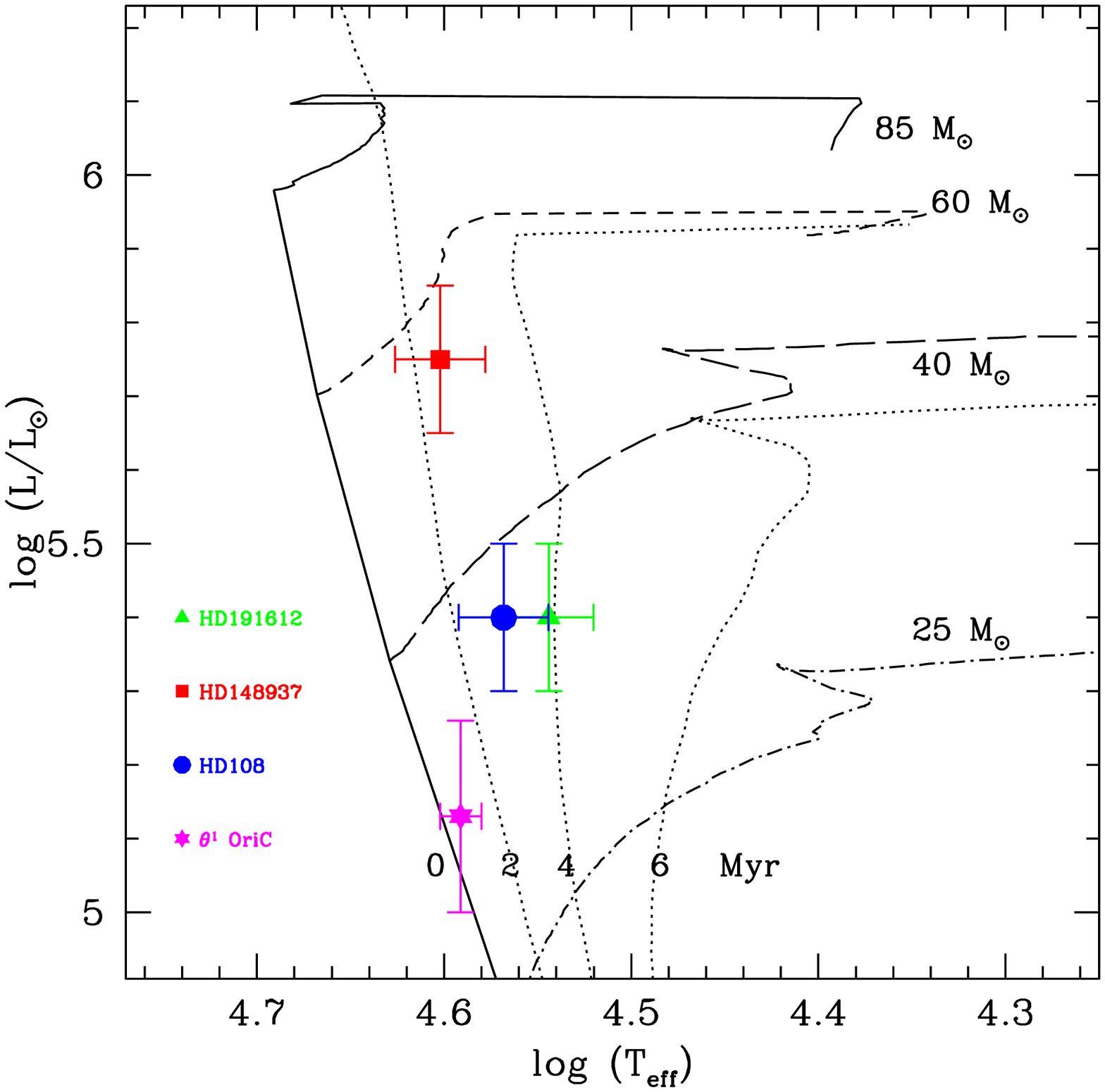}
  \caption{HR diagram with Geneva isochrones and evolutionary tracks from \citet{mey05}.}
  \label{fig:diaghr}
\end{minipage}
\begin{minipage}{8cm}
  \centering
  \includegraphics[width=7.5cm]{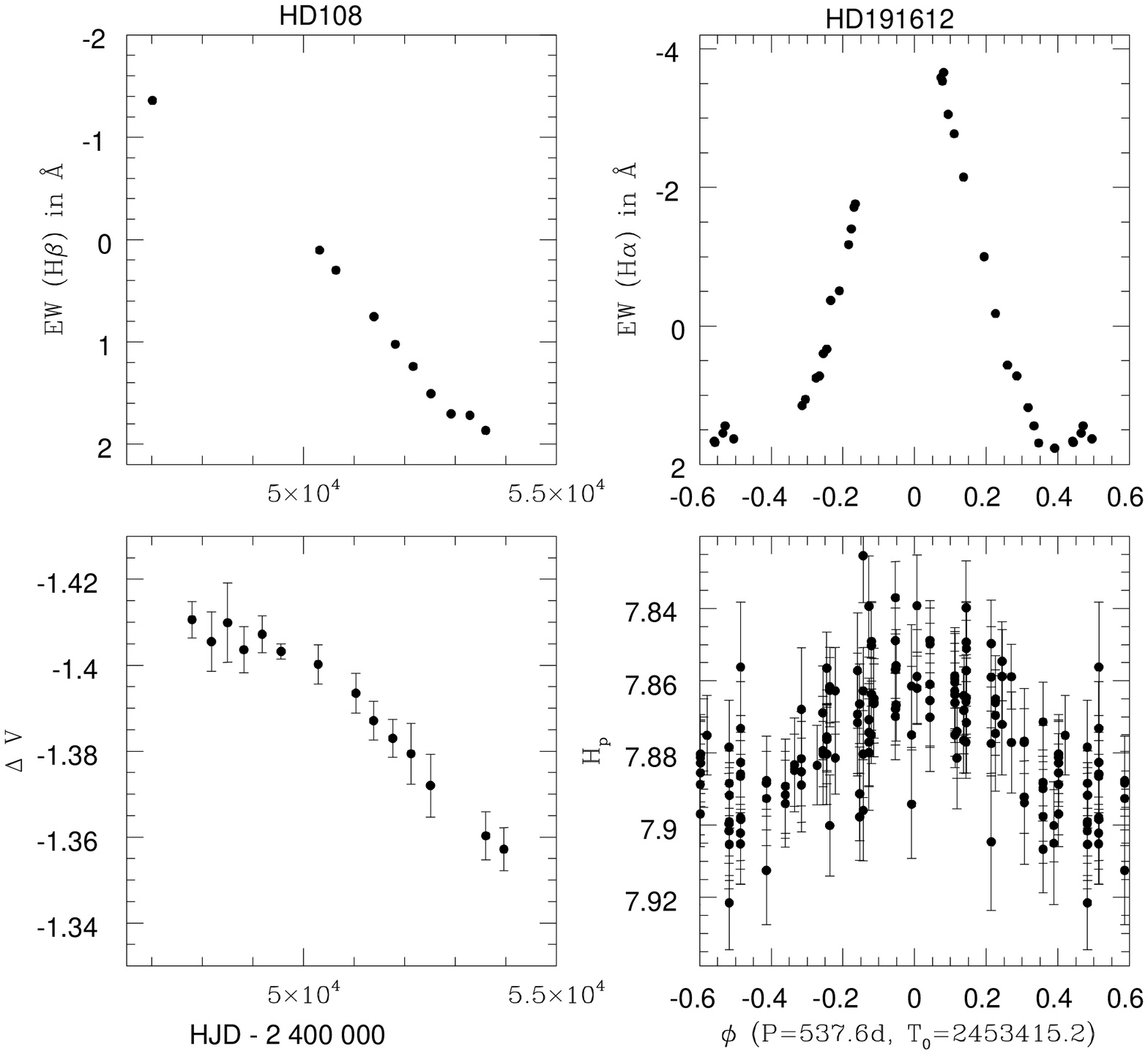}
  \caption{Photometric and EW changes of \hda\ (left, data from \citealt{naz06} and \citealt{bar07}) and \hdb\ (right, data from 
  Hipparcos archives and \citealt{naz07}). Negative EWs correspond to emission features.}
  \label{fig:photom}
\end{minipage}
\end{figure}

Fig. \ref{fig:diaghr} presents the positions of the stars in the HR 
diagram: two stars are close together, suggesting that they share 
similar properties, whereas \hdc\ appears clearly more massive and 
more luminous and \thet\ appears only slightly less massive and less 
luminous. Note that for the luminosity estimates, the distances to the
Of?p stars are based on their supposed membership in OB associations
(2.51\,kpc, Cas~OB5 for \hda; 2.29\,kpc, Cyg~OB3 for \hdb; and 1.38\,kpc,
Ara~OB1a for \hdc, see \citealt{hum78}), which may be uncertain.

The masses were estimated by several methods: orbital solutions for
binary objects, model-atmosphere fits ($M_{spec}=gR^2/G$), and positions 
in the HR diagram compared to predictions of evolutionary models. 
The values derived by the first and last methods are listed in 
Table \ref{tab:physpar}, with a superscript indicating the method used. 
The different determinations generally coincide very closely for a 
given star, except for \hdc\ where the HR-diagram mass is at 1$\sigma$ 
from the spectroscopic mass (82$\pm$33\,\msol), but still within the 
error bar. 

The ages of the stars were estimated by comparing their positions in 
the HR diagram to theoretical isochrones, which yields 2--4\,Myr for the 
Of?p objects and $\sim$1\,Myr for \thet. The ages of the associations 
to which the stars belong generally provide a more accurate estimate. 
For our objects, this method yields $<$1--2\,Myr for \thet\ \citep{hil97}, 
2--5\,Myr for \hdb\ \citep{mas95}, and 0--3\,Myr for \hdc\ \citep{vaz92}. 
\thet\ seems to be the youngest system of the sample, in agreement with 
its location in the Orion Nebula Cluster, but the uncertainties on the 
ages of the Of?p objects make a more detailed comparison difficult.

\begin{sidewaystable*}
\centering
  \setlength{\tabnotewidth}{0.5\columnwidth}
  \tablecols{9}
  \caption{Physical parameters of \thet\ and the Of?p stars. }
\label{tab:physpar}
 \begin{tabular}{lllllllll}
    \toprule
Star & $T_{eff}$& $\log \left( \frac{L}{L_{\odot}} \right)$ & $\log(g)$ & $\frac{R}{R_{\odot}}$ & $v\,\sin(i)$ & Age & $\frac{M}{M_{\odot}}$ & References\\
 &  (kK) &  & & & (\kms) & (Myr) & & \\
    \midrule
\thet & 39$\pm$1 & 5.13$\pm$0.13 & 4.1$\pm$0.1 & 8.1$\pm$1.3 & 24$\pm$3 & 0--2\tabnotemark{a} & $\sim$30\tabnotemark{a} & \citet{sim06}, this work\\
      &39.9+31.9 & 5.31+4.58     &             &            &          &             & 34.0+15.5?\tabnotemark{b}  & \citet{krau07}, this work\\
\hda  & 37$\pm$2 & 5.40$\pm$0.10 &3.75$\pm$0.10&12.3$\pm$2.1 & $\gtrsim$40 & 2--4\tabnotemark{a} & $\sim$35\tabnotemark{a} & this work\\
\hdc  & 41$\pm$2 & 5.75$\pm$0.10 & 4.0$\pm$0.1 &15.0$\pm$2.5 & $\gtrsim$45 & 2--4\tabnotemark{a} & $\sim$55\tabnotemark{a} & \citet{naz08}, this work\\
\hdb  & 35$\pm$1 & 5.4           & 3.5$\pm$0.1 & 14.5       & $\gtrsim$45 & 2--4\tabnotemark{a} & 30+15?\tabnotemark{b}/$\sim$35\tabnotemark{a} & \citet{how07}, this work\\
    \bottomrule
\tabnotetext{a}{Estimated from the position of the star in the HR diagram.}
\tabnotetext{b}{Estimated from the orbital solution.}
  \end{tabular}
\end{sidewaystable*}

\subsubsection{Evolutionary Status: Peculiar Abundances and Circumstellar Material}

Model atmosphere fits are also able to reveal the abudance pattern.
For \hda\ and \hdc, they unveil a clear overabundance of nitrogen: for 
\hda, N/H is $6\times10^{-4}$, i.e. a value about 9 times solar (this 
work) while it is $3\times10^{-4}$, or about 4 times the solar value, for 
\hdc\ \citep{naz08}. A comparison of line strengths in the spectra of the 
Of?p stars suggests a similar overabundance for \hdb. This enrichment, 
together with the presence of common spectral features, points towards 
similarities between the Of?p stars and Ofpe/WN9 objects \citep{wal03},
suggesting that these stars might be slightly evolved objects.

In this context, it is interesting to note that \hdc\ is surrounded by 
a circumstellar nebula: the bipolar nebula NGC~6164-6165. The northwest 
lobe, NGC\,6164, is receding while the southeast lobe, NGC\,6165, is 
approaching; this suggests an expansion, with a projected velocity of 30\,\kms\ \citep{lei87}. Moreover, the nebula displays anomalous chemical abundances, 
and it was therefore suggested to have been formed through an eruption 
of the Of?p central star \citep{lei87,duf88}. Evidence of this process 
can be found in the similar nitrogen overabundances of the nebular and 
stellar data \citep[factor of 4--5 compared to the solar value; 
][]{duf88,naz08}. It is possible that the lower amplitude of the 
line-profile variability for \hdc\ arises from a relaxation of the 
system following a Luminous-Blue-Variable-like eruption. 

\section{Photometric variability}

Hipparcos broad-band photometry of \hdb\ exhibits a clear modulation with a 
period of $\sim$538\,d \citep{koe02,naz04}. The spectroscopic and photometric
variations are clearly in phase: when the emission lines are weakest and the 
star displays an O8 spectral type, the star is fainter; when the emissions 
are maximum and the spectral type is earlier, then the star is brighter 
\citep{wal04}.

Variable photometry was also reported for \hda\ \citep{bar99, bar07}. 
In fact, the brightness of the star clearly decreased in recent years 
($\Delta$V=0.06mag). No significant color variation was found, but the 
declining luminosity clearly correlates with stronger \hei\ absorptions 
and weaker H emission lines in the visible spectrum of the star: the 
photometric variability of \hda\ and \hdb\ thus appears very similar 
(Fig. \ref{fig:photom}). 

As \hda, \hdc\ was included in the ``New Catalogue of Suspected Variable 
Stars'' \citep{kuk81} but its variability status is actually not ascertained. 
On the one hand, short-term variability was observed by \citet{bal92}: 
the star dimmed by 0.01~mag over a few weeks. On the other hand, 
\citet{van89} consider the star to have a constant luminosity in $V$, 
though with possible color changes ($\sim$0.002~mag); a larger dispersion 
of the magnitude might have been observed during some observing runs. 
The same authors further suggested that \hdc\ might have been bluer 
and brighter in the late eighties than in 1960. Finally, \hdc\ was 
classified as a possible candidate S\,Dor variable, but with only weak
indications for that status \citep{van01}.

No thorough photometric variability analysis has been made for \thet. 
However, it is also included in the ``New Catalogue of Suspected 
Variable Stars'' \citep{kuk81} and Hipparcos data display some (apparently) stochastic variability with an amplitude of about 0.14~mag (between 4.56 and 4.70~mag, \citealt{per97}). \citet{sim06} suggested that the 15d variability 
of the `photospheric'-line EWs could be explained by dilution by a varying 
continuum. However, other scenarios have been proposed: excess absorption 
due to corotating clouds when the star is seen pole-on \citep{sta96}, or 
excess emission due to infalling material when the star is seen equator-on 
\citep{smi05,wad06}. Photometric changes linked to the 15d period thus 
remain to be identified.

\section{X-Ray Properties}

Several high-resolution X-ray spectra of \thet\ were obtained with the Chandra 
Observatory \citep{sch00,gag05}. These data revealed that \thet\ displays a 
very hard X-ray spectrum, with a clear overluminosity 
($\log[L_X/L_{BOL}]=-6.0$, to be compared with the `canonical' value 
of $-$6.9 from \citealt{san06}). The spectrum is mainly thermal 
in nature, with two components of temperatures 0.7~keV and 2.5--3~keV 
\citep{gag05}. The hottest plasma clearly dominates the spectrum 
(Fig. \ref{fig:dem}). In addition, the X-ray lines appear very narrow 
(FWHM$\sim$600\kms, i.e. much less than the wind terminal velocity), 
as expected for a magnetic oblique rotator model \citep{bab97,gag05}. 
Finally, the X-ray emission is not constant, but varies in phase with 
the H line emissions of the visible spectrum. 

\begin{figure}
  \centering
  \caption{Differential Emission Measure of \thet\ (left, from \citealt{zhe07}) and \hdb\ (right, from \citealt{naz07}). }
  \label{fig:dem}
\end{figure}

The \xmm\ observations of Of?p stars detected large overluminosities 
($\log[L_X/L_{BOL}]=-6.0$ to $-6.2$), but they also unveiled crucial 
differences from \thet\ \citep{naz04b,naz07,naz08}. First, the X-ray 
spectra of the Of?p objects are rather soft. In the best fits, two 
temperatures are found but the lower one (0.2--0.3~keV) clearly 
dominates (Fig. \ref{fig:dem}). In fact, the component at higher 
temperature (1--3~keV) accounts for only 30\% of the unabsorbed
flux: it can thus not explain the overluminosity by itself. Second, 
the X-ray lines are rather broad (FWHM$\sim$1800\kms, a value similar 
to the wind terminal velocity), as is found for `normal' O-type stars.

An intriguing characteristic of \hdb\ must be noted: as for \thet, its
 X-ray flux decreased (by 40\%) as the emission lines declined in the 
visible spectrum. However, since all observations were taken during the 
same 538d cycle, it is not yet known whether this variability is 
phase-locked and if so, with which period (the period of the binary, 
1542d, or the period of the line profile variations, 538d).

\section{Magnetic Observations}

\citet{don02} and \citet{don06} reported the detection of
magnetic fields in \thet\ and \hdb, respectively, through the analysis 
of the Zeeman signature in several lines of the visible spectrum. 
With multiple observations of this signature throughout the 15d cycle, 
it clearly seems that \thet\ is indeed a magnetic oblique rotator 
as had been suggested before. In this model, the modulation observed 
in the visible and X-ray domains comes from the viewing angle of 
the magnetically confined disk, which changes with the stellar rotation, 
i.e. stronger emissions are detected when the disk is seen face-on 
\citep[for a more detailed modeling see][]{smi05}. \citet{don06} then 
proposed \hdb\ to be an evolved version of \thet. In this case, the 538d 
timescale would be the rotation period of the star, which would have 
been braked as a result of the intense magnetic field. Using also
spectropolarimetric data, \citet{hub08} recently reported the detection 
of a magnetic field for \hdc. Additional spectropolarimetric measurements 
of \hdb\ and \hdc\ are now needed to check if the observed magnetic field
follows the phase-locked evolution expected for magnetic oblique rotators.

\section{Summary and Conclusions}

The unusual Of?p stars and \thet\ share several similarities but display 
also intriguing differences (see Table \ref{tab:compa}). 

In the visible domain, typical lines of O-type spectra, several being 
in emission or having P Cygni profiles, are observed for all stars, 
but the Of?p stars present a rare feature in addition: strong
\ciii\,\lo\,4650 emission, the exact origin of which is still unknown. 
Spectral variability is detected for all objects, but the details vary 
from one case to another. The common characteristic is the presence of 
variations of the H and \heii\,\lo\,4686 lines. The \hei\ and \ciii\ 
lines also show large profile changes, but only for \hda\ and \hdb; 
for \thet, photospheric lines also vary, but at a low level, which 
might be due to variations in the continuum level; and the observed 
variability of \hdc\ has the lowest intensity of all. These line-profile 
variations appear recurrent in each case, but with very different 
timescales: 7d for \hdc, 15d for \thet, 538d for \hdb, and 55~yr for 
\hda. These periods are apparently not linked to orbital motion: 
the two known binaries (\hdb\ and \thet) present much longer orbital 
periods. However, the periods are related to changes seen in X-rays (for 
\thet\ and possibly \hdb) or in broad-band photometry (for \hdb\ and \hda). 
For the latter two objects, the spectral type appears earlier (due 
to filling in of the \hei\ lines) when the brightness is greater and 
the emission lines stronger. 

The physical parameters are also quite similar (Table \ref{tab:physpar}): 
low projected rotational velocities, temperatures of about 40~kK, radii 
and gravities favoring a main-sequence or giant classification. However, 
the more massive \hdc\ is surrounded by a circumstellar nebula and all 
the Of?p objects display nitrogen enrichment in their spectra: these 
stars are thus likely slightly more evolved than \thet.  

So far, magnetic fields have been detected in only a few O-type stars,
among them \thet, \hdc\ and \hdb. Without further observations, however, we 
can not exclude the presence of magnetism in \hda.

In the X-ray domain, the differences between the Of?p stars and \thet\ 
seem more pronounced. Although all present overluminosities compared to 
normal O-type stars, only \thet\ displays narrow lines and the hard spectrum 
expected for a magnetically confined wind.

In conclusion, \hdb\ (or the other Of?p stars) cannot `simply' be identical
to \thet\, since significant differences are clearly detected 
in their behaviors. On the other hand, their similarities should not be 
disregarded: for example, the unsual presence of a second peak in the 
differential emission measure of the Of?p high-energy spectrum could 
indicate the presence of a magnetically-confined wind, though it does not
dominate the X-ray properties. It is therefore probable that the true 
nature of these objects is actually dual: something similar to \thet\ 
plus a yet unknown phenomenon, perhaps related to the age difference. 
In this context, understanding the exact origin of the \ciii\ emission 
in the Of?p objects might be a crucial step.

\begin{table*}
\centering
  \setlength{\tabnotewidth}{0.5\columnwidth}
  \tablecols{5}
  \caption{Comparison of the properties of \thet\ and the Of?p stars.} \label{tab:compa}
 \begin{tabular}{lcccc}
    \toprule
 & \thet & \hda & \hdc & \hdb \\
    \midrule
Sp. Type & O7V & O4f?p$\leftrightarrow$O8.5fp & O5.5--6f?p & O6.5f?pe$\leftrightarrow$O8fp \\
HI& var& var& var& var\\
\hei & var& var& cst & var \\
\heii\,\lo\,4686 & var& var& var & var\\
\ciii\,\lo\,4650 & not pr.& var& cst & var\\
Line prof. Period & 15d& 55yr& 7d & 538d\\
Binary Period & 11--26yr& N?& N?& 1542d\\
Magn. Field & Y& ?& Y& Y\\
N enrichment & N& Y& Y & Y\\
Photometry & var?& L$\downarrow$ when H$\downarrow$ & cst? & L$\downarrow$ when H$\downarrow$\\
X-ray excess & Y& Y& Y& Y\\
X-ray prop. & hard, narrow& \multicolumn{3}{c}{soft, broad lines} \\
    \bottomrule
\end{tabular}
\end{table*}

\acknowledgments
The authors thank Professors Stahl and Zhekov, as well as the editors of
A\&A, MNRAS, and IBVS, for permitting the use of published images. They 
also thank Ph. Eenens for providing an echelle spectrum of HD108 and the 
referee for providing helpful suggestions to improve the quality 
of the paper. YN acknowledges support from the Fonds National de la 
Recherche Scientifique (Belgium), the PRODEX XMM and Integral contracts, 
and the visitor's program of the STScI. STScI participation in the Small 
and Moderate Aperture Research Telescope System (SMARTS) Consortium at 
CTIO is funded by the Director's Discretionary Research Fund. STScI is 
operated by the Association of Universities for Research in Astronomy, 
Inc., under NASA contract NAS5-26555. FM thanks John Hillier for constant 
help with his code CMFGEN. \\


\begin{thebibliography}
\bibitem[Andrillat et al.(1973)]{and73} Andrillat, Y., Fehrenbach, C., Swings, P., \& Vreux, J.~M.\ 1973, \aap, 29, 171 
\bibitem[Aslanov \& Barannikov(1989)]{asl89} Aslanov, A.~A., \& Barannikov, A.~A.\ 1989, Soviet Astronomy Letters, 15, 316 
\bibitem[Babel \& Montmerle(1997)]{bab97} Babel, J., \& Montmerle, T.\ 1997, \apjl, 485, L29 
\bibitem[Balona\ (1992)]{bal92} Balona, L.A. 1992, MNRAS, 254, 404
\bibitem[Barannikov(1999)]{bar99} Barannikov A.A.,1999, Astron. Let. 25, 169
\bibitem[Barannikov(2007)]{bar07} Barannikov, A.~A.\ 2007, IBVS, 5756, 1
\bibitem[Bekenstein(1976)]{bek76} Bekenstein, J.~D.\ 1976, \apj, 210, 544 
\bibitem[Conti(1972)]{con72} Conti, P.~S.\ 1972, \apjl, 174, L79 
\bibitem[Conti et al.\ (1977)]{con77} Conti, P.S., Garmany, C.D., \& Hutchings, J.B.  1977, ApJ, 215, 561
\bibitem[Donati et al.(2002)]{don02} Donati, J.-F., Babel, J., Harries, T.~J., Howarth, I.~D., Petit, P., \& Semel, M.\ 2002, \mnras, 333, 55 
\bibitem[Donati et al.\ (2006)]{don06} Donati, J.-F., Howarth, I.D., Bouret, J.-C., Petit, P., Catala, C., \& Landstreet, J.  2006, MNRAS, 365, L6
\bibitem[Dufour et al.\ (1988)]{duf88} Dufour, R.J., Parker, R.A.R., \& Henize, K.G. 1988, ApJ, 327, 859
\bibitem[Evans et al.(2004)]{eva04} Evans, C.~J., Howarth, I.~D., Irwin, M.~J., Burnley, A.~W., \& Harries, T.~J.\ 2004, \mnras, 353, 601 
\bibitem[Gagn\'e et al.(2005)]{gag05} Gagn\'e M., Oksala M.E., Cohen D.H., et al.\ 2005, ApJ, 628, 986
\bibitem[Garmany et al.\ (1980)]{gar80} Garmany, C.D., Conti, P.S., \& Massey, P. 1980, ApJ, 242, 1063
\bibitem[Heydari-Malayeri \& Melnick\ (1992)]{hey92} Heydari-Malayeri, M., \& Melnick, J. 1992, A\&A, 258, L13
\bibitem[Hubrig et al.(2008)]{hub08} Hubrig, S., et al. 2008, A\&A, submitted
\bibitem[Hillenbrand(1997)]{hil97} Hillenbrand, L.~A.\ 1997, \aj, 113, 1733 
\bibitem[Howarth et al.\ (2007)]{how07} Howarth, I.D., Walborn, N.R., Lennon, D.J., et al. 2007, MNRAS, 381, 433
\bibitem[Humphreys(1978)]{hum78} Humphreys, R.~M.\ 1978, \apjs, 38, 309 
\bibitem[Hutchings\ (1975)]{hut75} Hutchings J.B., 1975, ApJ, 200, 122
\bibitem[Kukarkin et al.(1981)]{kuk81} Kukarkin, B.~V., et 
al.\ 1981, Moscow, Acad.~of Sciences USSR Shternberg,1951 (1981), 0 \bibitem[Koen \& Eyer (2002)]{koe02} Koen C., \& Eyer L. 2002, MNRAS, 331, 45
\bibitem[Kraus et al.(2007)]{krau07} Kraus, S., et al.\ 2007, \aap, 466, 649 
\bibitem[Leitherer \& Chavarria-K.(1987)]{lei87} Leitherer, C., \& Chavarria-K., C.\ 1987, \aap, 175, 208 
\bibitem[Massey et al.(1995)]{mas95} Massey, P., Johnson, 
K.~E., \& Degioia-Eastwood, K.\ 1995, \apj, 454, 151
\bibitem[Massey \& Duffy\ (2001)]{mas01} Massey, P., \& Duffy, A.S. 2001, ApJ, 550, 713
\bibitem[Menten et al.(2007)]{men07} Menten, K.~M., Reid, M.~J., Forbrich, J., \& Brunthaler, A.\ 2007, \aap, 474, 515 
\bibitem[Meynet \& Maeder(2005)]{mey05} Meynet, G., \& Maeder, A.\ 2005, \aap, 429, 581 
\bibitem[Naz\'e\ (2004)]{naz04} Naz\'e Y. 2004, PhD thesis, Universit\'e de Li\`ege
\bibitem[Naz\'e et al.\ (2001)]{naz01} Naz\'e, Y., Vreux, J.-M., Rauw, G. 2001, A\&A 372, 195
\bibitem[Naz\'e et al.\ (2004)]{naz04b} Naz\'e, Y.,  Rauw, G., Vreux, J.-M., \& De Becker, M.\ 2004, A\&A, 417, 667
\bibitem[Naz\'e et al.\ (2006)]{naz06} Naz\'e, Y., Barbieri, C., Segafredo, A., Rauw, G., \& De Becker, M.\ 2006, IBVS, 5693
\bibitem[Naz\'e et al.\ (2007)]{naz07} Naz\'e, Y.,  Rauw, G., Pollock, A.M.T., Walborn, N.R., \& Howarth, I.D.\ 2007, MNRAS, 375, 145
\bibitem[Naz\'e et al.\ (2008)]{naz08} Naz\'e, Y.,  Walborn, N.R., Rauw, G., Martins, F., Pollock, A.M.T., \& Bond, H.E.\ 2008, AJ, 135, 1946
\bibitem[O'Dell(2001)]{ode01} O'Dell, C.~R.\ 2001, \pasp, 113, 29 
\bibitem[Patience et al.(2008)]{pat08} Patience, J., Zavala, R.~T., Prato, L., Franz, O., Wasserman, L., Tycner, C., Hutter, D.~J., \& Hummel, C.~A.\ 2008, \apjl, 674, L97 
\bibitem[Perryman \& ESA(1997)]{per97} Perryman, M.~A.~C., \& ESA 1997, ESA Special Publication, 1200
\bibitem[Sana et al.\ (2006)]{san06} Sana H.,  Rauw G., Naz\'e Y., Gosset E., \& Vreux J.-M. 2006, MNRAS, 372, 661
\bibitem[Schulz et al.(2000)]{sch00} Schulz, N.S., Canizares, C.R., Huenemoerder D., \& Lee, J.C. 2000, ApJ, 545, L135
\bibitem[Sim{\'o}n-D{\'{\i}}az et al.(2006)]{sim06} Sim{\'o}n-D{\'{\i}}az, S., Herrero, A., Esteban, C., \& Najarro, F.\ 2006, \aap, 448, 351 
\bibitem[Sim\'on-D\'{\i}az \& Herrero\ (2007)]{sim07} Sim\'on-D\'{\i}az, S., \& Herrero, A. 2007, A\&A, 468, 1063
\bibitem[Smith \& Fullerton(2005)]{smi05} Smith, M.~A., \& Fullerton, A.~W.\ 2005, \pasp, 117, 13 
\bibitem[Stahl et al.(1993)]{sta93} Stahl, O., Wolf, B., G\"ang, T., Gummersbach, C.A., Kaufer, A., \& Kovacs, J. 1993, A\&A, 274, L29
\bibitem[Stahl et al.(1996)]{sta96} Stahl, O., et al.\ 1996, \aap, 312, 539 
\bibitem[Stahl et al.(2008)]{sta08} Stahl, O., Wade, G., Petit, V., Stober, B., \& Schanne, L. 2008, A\&A, in press (arXiv:0805.0701)
\bibitem[Underhill(1994)]{und94} Underhill, A.~B.\ 1994, \apj, 420, 869 
\bibitem[van Genderen\ (2001)]{van01} van Genderen, A.M.  2001, A\&A, 366, 508 
\bibitem[van Genderen et al.\ (1989)]{van89} van Genderen, A.M., Bovenschen, H., Engelsman, E.C., et al. 1989, A\&AS, 79, 263
\bibitem[Vazquez \& Feinstein(1992)]{vaz92} Vazquez, R.~A., 
\& Feinstein, A.\ 1992, \aaps, 92, 863 
\bibitem[Vreux \& Conti(1979)]{vre79} Vreux, J.-M., \& Conti, P.S.\ 1979, \apj, 228, 220 
\bibitem[Wade et al.(2006)]{wad06} Wade, G.~A., Fullerton, A.~W., Donati, J.-F., Landstreet, J.~D., Petit, P., \& Strasser, S.\ 2006, \aap, 451, 195 
\bibitem[Walborn(1972)]{wal72} Walborn, N.R. 1972, AJ, 77, 312
\bibitem[Walborn(1973)]{wal73} Walborn N.R. 1973, AJ, 78, 1067
\bibitem[Walborn(1981)]{wal81} Walborn, N.R.\ 1981, \apjl, 243, L37 
\bibitem[Walborn \& Nichols(1994)]{wal94} Walborn, N.R., \& Nichols, J.S.\ 1994, \apjl, 425, L29 
\bibitem[Walborn et al.\ (2000)]{wal00} Walborn, N.R., Lennon, D.J., Heap, S.R., Lindler, D.J., Smith, L.J., Evans, C.J., 
\& Parker, J.Wm. 2000, PASP, 112, 1243
\bibitem[Walborn et al.\ (2003)]{wal03} Walborn, N.R., Howarth, I.D., Herrero, A., \& Lennon, D.J. 2003, ApJ, 588, 1025
\bibitem[Walborn et al.\ (2004)]{wal04} Walborn, N.R., Howarth, I.D., Rauw, G., et al. 2004, ApJ, 617, L61 
\bibitem[Zhekov \& Palla\ (2007)]{zhe07} Zhekov, S.A., \& Palla, F. 2007, MNRAS, 382, 1124

\end{thebibliography}
\end{document}